\documentclass[11pt,preprint]{aastex}
\usepackage{amsmath}

\slugcomment{\today}

\def\VR{{\ensuremath{V\!R}}}

\def\sm{SuperMACHO}

\def\dsct{$\delta$-Scutis}
\def\dsctsing{$\delta$-Scuti}

\begin{document}

\title{High Amplitude \dsct~in the Large Magellanic Cloud}

\shorttitle{LMC \dsct}
\shortauthors{Garg et al.}

\author{A. Garg\altaffilmark{1,2}, 
K. H. Cook\altaffilmark{1,2}, 
S. Nikolaev\altaffilmark{1},
M. E. Huber\altaffilmark{2,3},
A. Rest\altaffilmark{4,5,6}, 
A. C. Becker\altaffilmark{7},
P. Challis\altaffilmark{2,8},
A. Clocchiatti\altaffilmark{9},
G. Miknaitis\altaffilmark{2,11}, 
D. Minniti\altaffilmark{9,10}, 
L. Morelli\altaffilmark{12},
K. Olsen\altaffilmark{5,13}, 
J. L. Prieto\altaffilmark{2,14},
N. B. Suntzeff\altaffilmark{5,15,16}, 
D. L. Welch\altaffilmark{2,17},
W. M. Wood-Vasey\altaffilmark{2,18}
}

\altaffiltext{1}{Lawrence Livermore National Laboratory, Institute of Geophysics and Planetary Physics, 7000 East Ave., Livermore, CA 94550}

\altaffiltext{2}{Visiting Astronomer, Cerro Tololo Inter-American
Observatory, National Optical Astronomy Observatory, which is operated
by the Association of Universities for Research in Astronomy,
Inc. (AURA) under cooperative agreement with the National Science
Foundation}

\altaffiltext{3}{Johns Hopkins University, Baltimore, MD 21218}
\altaffiltext{4}{Dept. of Physics, Harvard University, 17 Oxford Street, Cambridge, MA 02138}
\altaffiltext{5}{Cerro Tololo Inter-American Observatory, National Optical Astronomy Observatory (CTIO/NOAO), Colina el Pino S/N, La Serena, Chile}
\altaffiltext{6}{Goldberg Fellow}
\altaffiltext{7}{Dept. of Astronomy, University of Washington, Box 351580, Seattle, WA 98195}
\altaffiltext{8}{Harvard-Smithsonian Center for Astrophysics, 60 Garden St., Cambridge, MA 02138} 
\altaffiltext{9}{Dept. of Astronomy, Pontificia Universidad Cat\'olica de Chile, Casilla 306, Santiago 22, Chile}
\altaffiltext{10}{Vatican Observatory, V00120 Vatican City State, Italy}
\altaffiltext{11}{Center for Neighborhood Technology, 2125 W. North Ave., Chicago IL 60647}
\altaffiltext{12}{Dipartimento di Astronomia, Universit\`a di Padova, vicolo dell'Osservatorio~3, I-35122 Padova, Italy}
\altaffiltext{13}{National Optical Astronomy Observatory, 950 N. Cherry Ave., Tucson, AZ 85719}
\altaffiltext{14}{Dept. of Astronomy, Ohio State University, 140 West 18th Ave., Columbus, OH 43210}
\altaffiltext{15}{Dept. of Physics and Astronomy, Texas A\&M University, College Station, TX 77843-4242}
\altaffiltext{16}{Mitchell Institute for Fundamental Physics, Texas A\&M University, College Station, TX 77843-4242}
\altaffiltext{17}{Dept. of Physics and Astronomy, McMaster University, Hamilton, Ontario, L8S 4M1, Canada}
\altaffiltext{18}{Dept. of Physics and Astronomy, University of Pittsburgh, 3951 O'Hara St., Pittsburgh, PA 15260}

\begin{abstract}
We present 2323 High-Amplitude \dsctsing~(HADS) candidates discovered in
the Large Magellanic Cloud (LMC) by the SuperMACHO survey
\citep{Rest05}.  Frequency analyses of these candidates reveal that
several are multimode pulsators, including 119 whose largest amplitude
of pulsation is in the fundamental (F) mode and 19 whose largest
amplitude of pulsation is in the first overtone (FO) mode.  Using
Fourier decomposition of the HADS light curves, we find that the
period-luminosity (PL) relation defined by the FO pulsators does not
show a clear separation from the PL-relation defined by the F
pulsators.  This differs from other instability strip pulsators such
as type c RR~Lyrae.  We also present evidence for a larger amplitude,
subluminous population of HADS similar to that observed in Fornax
\citep{Poretti08}.

\end{abstract}

\keywords{surveys---Magellanic Clouds---{\it Facilities:} \facility{Blanco ()}}

\section{Introduction}
\dsctsing~variables populate the region of the Hertzprung-Russell diagram
where the instability strip meets the main sequence.  The
high-amplitude variables are generally believed to be pulsating
primarily in radial modes, whereas \dsct~with smaller amplitudes are
believed to have many non-radial modes of pulsation.  \citet{Breger}
provides a thorough review of the theoretical models describing
\dsctsing~pulsation.  Recent space-based observations from the CoRoT
telescope have begun to reveal the rich complexity of \dsctsing~pulsation
modes \citep{Poretti09}.

Despite their multimode nature, HADS have been shown to define a
period-luminosity relationship, allowing for their use as standard
candles \citep[and references therein]{mcnamaraLMC,Poretti08}.  Until
recently, observations of large sets of HADS have been limited due to
these stars' intrinsic faintness and short periods.  The majority of
known HADS have been found within the Milky Way
(e.g. \citealt{machoblg}, \citealt{MACHOblgdsct}, and
\citealt{Pigulski06}).  More recent work has revealed 90 \dsct~(or
SX~Phoenicis stars, Population II \dsct) in Fornax \citep[hereafter
P08]{Poretti08}.  Using a subset of the MACHO project data,
\citet{machodsct} finds 101 \dsct~in the LMC.  \citet{mcnamaraLMC}
report 24 \dsct~in the LMC using the OGLE-II data set.  Their work
also provides a summary of HADS detected by ground-based surveys.

In this paper we present analyses of a large set of \dsct~discovered
by the SuperMACHO survey of the LMC.  In Section~\ref{sec:SM} we
provide an overview of the survey and data reduction.  In
Section~\ref{sec:data} we discuss our data and HADS selection
criteria, and we present our candidates.  We discuss our findings in
Section~\ref{sec:disc}.  We perform a frequency spectrum analysis of
the HADS candidates to identify multimode pulsators and present
evidence for a large set of FO pulsators.  We examine a subset of our
candidates having larger amplitudes.  We find evidence for an excess
population of faint sources, and discuss whether it is the subluminous
population observed in Fornax (P08). 

\section{The SuperMACHO Survey: Observations and Image Reduction}\label{sec:SM}

\subsection{Survey overview}
The SuperMACHO project is a five-year optical survey of the Large
Magellanic Cloud (LMC) aimed at detecting microlensing of LMC stars
\citep{Rest05}.  The goal of this survey is to determine the location
of the lens population responsible for the excess microlensing rate
observed toward the LMC by the MACHO project \citep[see][and
references therein]{Alcock00} and, thereby, better constrain the
fraction of MAssive Compact Halo Objects (MACHOs) in the Galactic
halo. The survey was conducted on the CTIO Blanco 4m telescope using a
custom $\VR$ broadband filter with the MOSAIC II wide-field
imager.  \citet{Garg07}, \citet{Miknaitis07},
\citet{Garg08}, and \citet{2008AIPC.1082..294R} provide a more
complete description of the survey observations and data reduction.
Here we provide a brief overview of key elements of the observations
and data reduction.  During each year of the survey, SuperMACHO
observed 68 LMC fields over $\sim$30 half-nights during dark and gray
time between the months of September -- January.  The nights and data
reduction pipeline were shared with the ESSENCE survey
\citep{Miknaitis07,WoodVasey07}.  The SuperMACHO survey was completed
in January 2006.

\subsection{Object selection}\label{section:objSel}
Variable objects are identified using the difference imaging algorithm
developed by Alard and Lupton \citep{Alard98,Alard00}.  The technique
is implemented using the ``High Order Transform of PSF and Template
Subtraction'' (HOTPANTS) software package.\footnote{See online manual
at http://www.astro.washington.edu/becker/hotpants.html.}  Excess or
negative flux, ``difference flux'', detections with a signal-to-noise
(S/N) $>$ 5 are matched between images.  Those coincident within
0.54'' (2~pixels) of each other from image-to-image are assumed to
belong to the same source, and the average position of these
detections is taken as the source position.  If there are at least
three detections of a source in any survey year, we identify the
source as a ``real'' variable object.

\subsection{Difference-image photometry} \label{section:diffphot}
Once we have have identified a source as variable, we perform
fixed-centroid, or {\it forced}, photometry at the detection centroid
to obtain a complete difference flux light curve
\citep{Garg07,Garg08}.  We measure the difference flux using a
modified version of DoPHOT \citep{Schechter93} that identifies sources
of negative flux.  Because there are few actual sources of difference
flux and many artifacts in any difference image, we cannot use the
difference image itself to determine the point spread function (PSF).
Instead we use the PSF determined for the science image prior to image
differencing.  We also force the centroid to be at the position
determined by the high S/N difference flux detections.

Because we neglect covariance terms when convolving the images for
differencing, we find that we underestimate the noise in the
difference image.  To empirically correct for this, we obtain the flux and its
uncertainty for a grid of positions across the difference image using
aperture photometry.  If the flux uncertainties reflect the Poisson
noise, then we would expect a histogram of ${\rm flux}/\delta {\rm flux}$ to
exhibit a gaussian distribution centered at 0.0 with a standard
deviation of 1.0.  For a typical image we find the standard deviation
to be closer to 1.5, so we adjust our flux uncertainties so that they
give a standard deviation of 1.0 which provides a more accurate
estimate of the uncertainty in the measured flux \citep{Garg08}.  We
also find that because it uses an analytical PSF model that does not
reflect additional structure in the PSF, DoPHOT further underestimates
the uncertainty in the measured flux by 0.01~mag \citep{Garg08}.  We
add this term in quadrature to the adjusted uncertainty returned by
our modified DoPHOT to obtain the uncertainties reported in this
paper.

\subsection{Multi-band imaging} \label{section:BIimaging}
In addition to the $\VR$ survey images, we also obtained a set of high
quality $B$- and $I$-band images.  We process these images using the
reduction pipeline described above.  We create $B$ and $I$ catalogs for these
images and generate $B-I$ color-magnitude diagrams (CMD) for each
field.  Though the MOSAIC II camera does not allow for simultaneous
imaging in multiple bands, these observations were typically made
within 5~minutes of each other.  The colors for even relatively short
period variables (1-2 hours) should be sufficiently accurate to
determine the rough position of sources in the CMD.

\section{Data and Results} \label{sec:data}

\subsection{Initial light-curve phasing}
After identifying all variable sources detected by SuperMACHO, we
phase their light curves to find periodic variables.  We perform light
curve phasing using both the SuperSmoother algorithm
\citep{1994PhDT........20R} and the CLEANest code
\citep{1995AJ....109.1889F}.  For SuperSmoother we use the period that
gives the SuperSmoother curve with the smallest residuals.  For
CLEANest we use the period corresponding to the highest power
frequency from the light curve's frequency spectrum.  For the initial
candidate selection, we take the shorter of the two periods to be the
light curve's period, $P$.  For the objects presented in this paper,
this is typically the period determined by CLEANest.  For all periodic
variables, we find the mean of the difference flux light curve (which
may be negative).  We add this to the template flux to determine the
mean magnitude of the variable object, $\VR$.

\subsection{Candidate selection} \label{section:Selection}

\begin{figure}
\centering
\includegraphics[angle=0,width=0.8\linewidth]{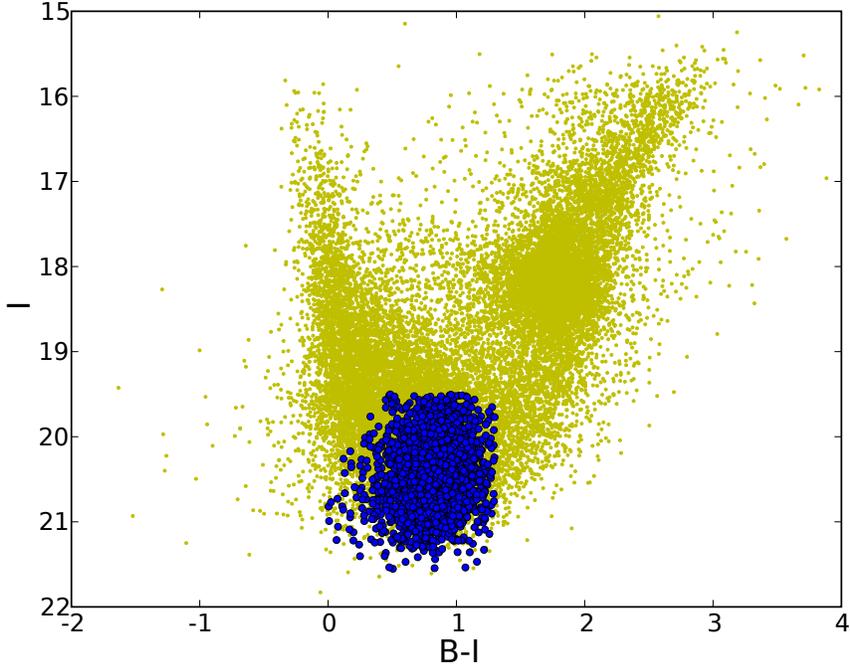}
\caption[]{$B-I$ color-magnitude diagram.  Yellow dots show all
star-type sources in the $B$ and $I$ catalog for a single amplifier in
field sm97.  The CMD is similar for all SuperMACHO fields.  The filled
blue circles show the final set of HADS candidates.  Candidate
selection criteria include the requirement that the source lie on the
main-sequence.}
\label{fig:CMDall}
\end{figure}

We present 2323 light curves of high-amplitude \dsct~(HADS) in the LMC
discovered by the SuperMACHO project.  We use the following criteria
to select these objects:

\begin{enumerate}
\item Using the values of $P$ and $\VR$ from the initial phasing, we
  create a period-luminosity (PL) diagram.  We find an overdensity of
  sources that lie roughly in the region expected for \dsct~in the
  LMC.  Based on the PL diagram, we define this region as $19.7 < \VR
  < 22.2$ and $0.045 < P < 0.145$.

\item We select a higher-amplitude subset of these candidates.  We choose
  only light curves for which the difference between the brightest and faintest
  data points is greater than 0.2~mag.

\item We cross-match the set of light curves against the $B$ and $I$
  catalogs described in Section~\ref{section:BIimaging} using a match
  radius of 0.27'' (1~pixel).  We remove candidates that do not have
  matches in both bands.  We also remove candidates that do not lie in
  the Main-Sequence portion of the CMD.  To ensure this, we require
  each candidate to have $I > 19.5$~mag and $0.0$~mag $< B-I <
  1.3$~mag.  Figure~\ref{fig:CMDall} shows the CMD for all \sm~sources
  falling on a single amplifier and the location of the
  \dsctsing~candidates.  After applying these selection criteria, we have
  4126 candidate sources.
  
\item We use the SigSpec \citep{SigSpec07} code to determine the
  frequency spectrum of the light curve (see
  Section~\ref{sec:SpecAn}).  In many cases, the primary frequency of
  variation found by SigSpec differs from that found during the
  initial phasing.  We select only light curves having a primary
  frequency of variation with spectral significance, $f_{\rm sig}$,
  greater than 5.475, which roughly corresponds to a signal-to-noise
  ratio of 4 \citep[see][]{SigSpec07}.  For the remainder of the
  analysis, we use the frequency associated with the highest amplitude
  of variability found by SigSpec such that $P = 1/f_{\rm SigSpec}$.
  Using the newly determined periods, we select only candidates with
  $0.045$~days $< P < 0.115$~days.\footnote{The shorter upper
  threshold on the period is based on the apparent region of
  \dsctsing~overdensity in the updated PL diagram.}

\item We perform a fourth order Fourier decomposition of the light
  curves (see Section~\ref{sec:Fourier}).  We use the coeffecient of
  the zero$^{th}$ order term, $a_0$, as the average stellar magnitude
  and keep candidates with $19.8 < a_0 < 22.2$.

\item We determine an overall amplitude using the model curve
  described by the fourth order Fourier fit.  We define the overall
  amplitude, $\Delta$, as the difference between the minimum and
  maximum points in the model curve, or the peak-to-trough amplitude.
  We select only those candidates with a $\Delta$ greater than
  0.2~mag.  This yields a final set of 2323 HADS candidates.  The
  majority of the candidates eliminated after applying the CMD
  criteria are eliminated by this amplitude cut.

\end{enumerate}

\begin{figure}
\centering
\includegraphics[angle=0,width=0.6\linewidth]{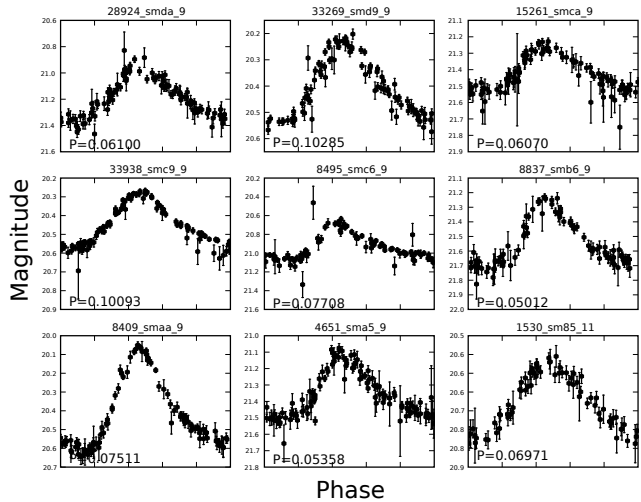}
\caption[]{Light curves for a selection of \dsctsing~candidates.}
\label{fig:LCs}
\end{figure}

To determine the efficacy of the above selection criteria, we have
inspected the light curves by eye.  We estimate that $<$2\% are
contaminants or poorly phased.  To retain objective selection
criteria, however, we leave them in the final set.
Table~\ref{tab:dScuti} gives the position, light-curve
characteristics, and color information for all the HADS candidates.
Figure~\ref{fig:LCs} shows a selection of candidate light curves.

\subsection{Comparison with other LMC \dsctsing~sets}

We compare our set of \dsct~to sets obtained from the OGLE-II
\citep{OGLEII} and MACHO surveys \citep{Alcock00}.
\citet{mcnamaraLMC} analyze a catalog of non-RR~Lyrae pulsating
variables accompanying the main \citet{OGLEII} catalog to find
evidence of 24 \dsctsing~candidates in the LMC.  We cross-match the
\citet{mcnamaraLMC} catalog against the set of {\it all} SuperMACHO
variables, including those we do not classify as \dsct.  We find only
1 match within 2.0'' of the position reported in \citet{mcnamaraLMC}.
This object, with OGLE-II designation 050309.65-684327.6, also appears
in our candidate set as 7151\_smc8\_15 and lies 1.54'' from the
OGLE-II position.  \citet{machodsct} reports 101 HADS discovered in a
subset of the MACHO database.  Of these, 99 fall within the FOV
observed by SuperMACHO.  We note that though these objects fall within
the SuperMACHO FOV, several may lie within gaps between amplifiers or
close to saturated sources that have been masked.  We estimate that
these considerations eliminate roughly 5\% of our FOV.  We cross-match
the \citet{machodsct} catalog to the set of SuperMACHO variables and
find 37 matches within 2'' of the MACHO positions.  Of these, 17
appear in our set of \dsctsing~candidates.  A majority of the eliminated
matches are removed based on the third criterion described above which
requires candidates to lie on the main-sequence described by the $B-I$
CMD.  We find, however, that without this criterion many of the
candidates lie along the red branch and have light curves similar to
close eclipsing binaries.

We consider why we find so few matches to these catalogs.  Some of the
sources identified in these catalogs may not meet the variable object
selection criteria described in Section~\ref{section:objSel}.  We
find, however, that a cross-match to the set of Blazhko RR~Lyrae
identified in MACHO \citep{MACHORRL} yields matches within 2'' for
62\% of the MACHO sources in the SuperMACHO FOV.  A more thorough
investigation of the SuperMACHO variable source detection efficiency
will appear in Garg et al. (2010, in prep); however, the larger match
fraction to the MACHO RR~Lyrae catalog suggests that low detection
efficiency is unlikely to account for the entire discrepancy.

We also examine some of the template images at positions where
\dsct~detected in MACHO were not found in SuperMACHO.  We do not see
any sources in the templates within several arcseconds of these
positions.  This suggests that the discrepancy does not result from a
failure to classify these objects as variable but rather from a lack
of sources at these positions.  This may suggest differences between
our astrometric solutions, though we emphasize that we do find at
least one match within 2'' in each catalog.  If we allow for a much
larger match radius of 5'', we find 44 matches to \citet{machodsct}
and 3 to \citet{mcnamaraLMC}.  Of these matches there are no new
candidates that pass our selection criteria.  We note that the
apparent brightnesses of LMC \dsct~are very close to the detection
limit for both MACHO and OGLE-II.  Misclassification of these faint
sources would not be surprising in these catalogs.

In addition to the LMC HADS reported by MACHO and OGLE-II, HADS have
been reported by the OGLE-III collaboration \citep{OIII08,OIII09}.
Notably the OGLE-III PL diagram suggests that the HADS have longer
periods than those discussed in this work.  This will be an important
area for further investigation when the final catalog becomes
available.

\subsection{Fourier decomposition}\label{sec:Fourier}

Using the period determined by SigSpec, we perform a fourth order Fourier
decomposition of the light curve to obtain the coefficients in the series

\begin{equation}
\VR(t) = a_{o} + \sum_{n=1}^{4} a_{n} \cos\left(\frac{2 \pi n t}{P}\right) + b_{n} \sin\left(\frac{2 \pi n t}{P}\right)
\label{eqn:expansion}
\end{equation}

\noindent where $\VR(t)$ is the observed $\VR$ magnitude at time $t$,
$n$ counts each order in the decomposition, $a_{n}$ and $b_{n}$ are the
coefficients of each order determined by the fit, and $P$ is
the period.  We find that a fourth order expansion is sufficient to
capture the structure of the light curves.  

Using the Fourier coefficients we calculate several additional
parameters describing the light curves.  $A_n$ is the amplitude of
pulsation for each order such that $A_n^2 = a_n^2 + b_n^2$.  $r_{ij}$
is the ratio of amplitudes such that $r_{ij} = A_i/A_j$.  $\phi_{n}$
is the phase shift for each order such that $\phi_{n} =
atan(b_n/a_n)$. $\phi_{ij}$ is given by $\phi_{ij} = \phi_{i} - i
\phi_j$.  These parameters are given in Table~\ref{tab:dScuti_par} for
all candidates.

\section{Discussion}\label{sec:disc}

\subsection{Period-Luminosity diagram}\label{sec:PLD}

\begin{figure}
\centering
\includegraphics[angle=0,width=0.6\linewidth]{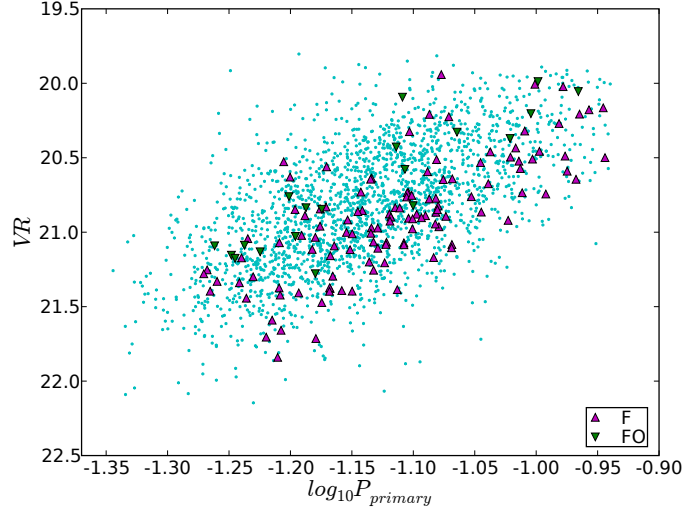}
\caption[]{Period-Luminosity diagram.  The cyan dots show all
\dsctsing~candidates. The green pentagons show multimode FO
pulsators, and the maroon diamonds show multimode F
pulsators (see Section~\ref{sec:SpecAn}).
}
\label{fig:dScutiPL_all}
\end{figure}

\begin{figure}
\centering
\plottwo{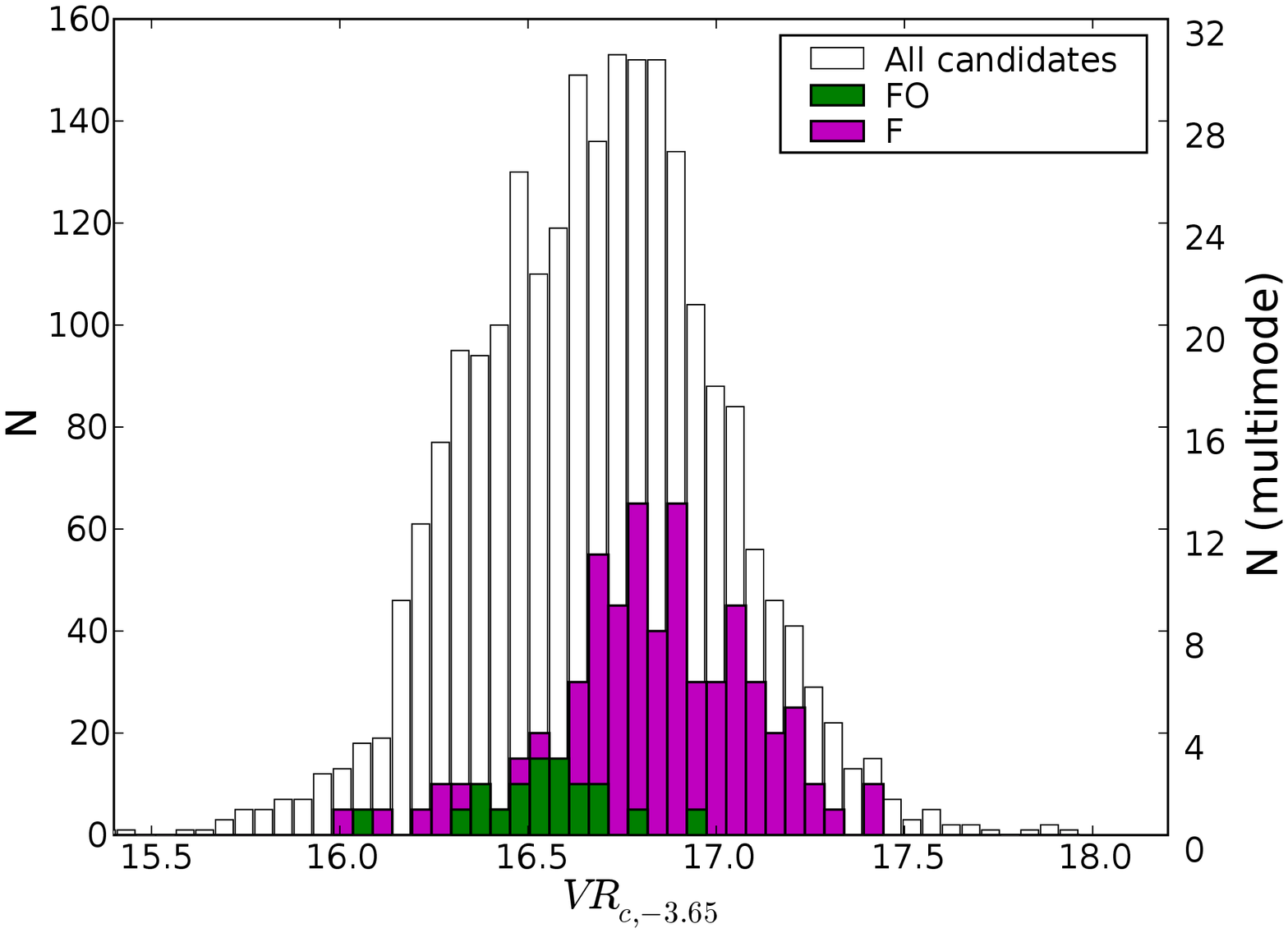}{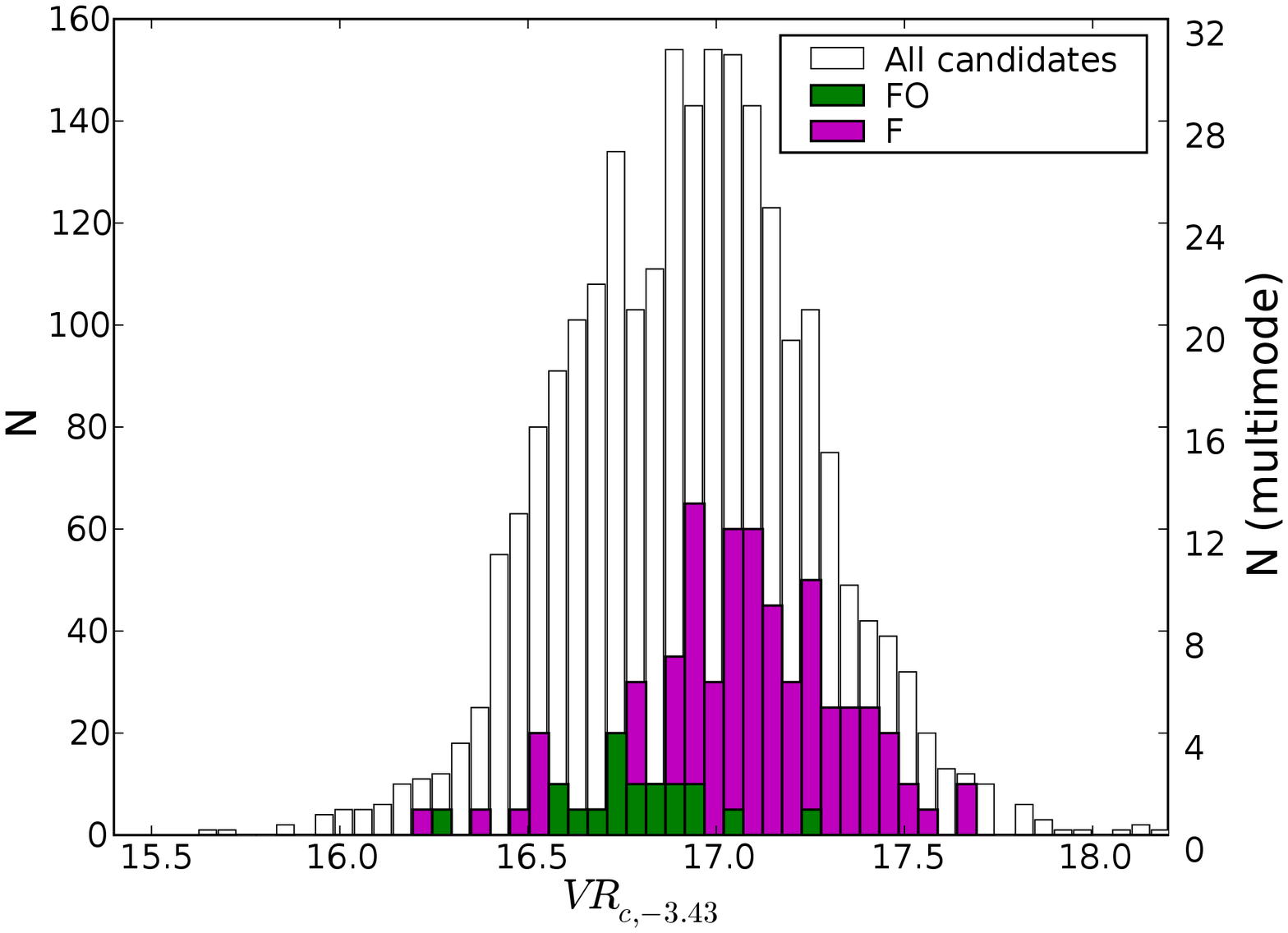}
\caption[]{Figure showing the histogram of PL-relation-corrected
luminosities, $\VR_c$.  We show histograms for the
PL-relation-corrected magnitudes of the candidates using both the
slope from P08 (left) and the slope independently determined from this data (right).
The double-mode F and FO pulsators are overplotted in maroon and green
respectively.  The histograms of the double-mode candidates are scaled
for visibility, and the scaling is shown on the right y-axis.  For
both slopes, we find the histogram is skewed toward brighter sources.
This excess of brighter sources is often interpreted as evidence
of overtone mode pulsators (see Section~\ref{sec:OT}).}
\label{fig:VRintHist}
\end{figure}

Figure~\ref{fig:dScutiPL_all} shows the period-luminosity (PL) diagram
of the final HADS candidates.  We note that the PL relation shows a
high degree of scatter.  This may indicate subgroups within the sample
such as the subluminous group observed by P08.  To identify such groups,
we create a histogram of the PL-relation-corrected luminosity, $\VR_c$,
of the candidates similar to P08 (Figure~\ref{fig:VRintHist}).  We
assume the slope of the $V$-band PL relation reported by P08 without
any transformation or metallicity correction.  This gives the
relation:

\begin{equation}
\VR_{c,-3.65} = 3.65 \log_{10}P + \VR
\label{eqn:vrint}
\end{equation}

\begin{figure}
\centering
\includegraphics[angle=0,width=0.6\linewidth]{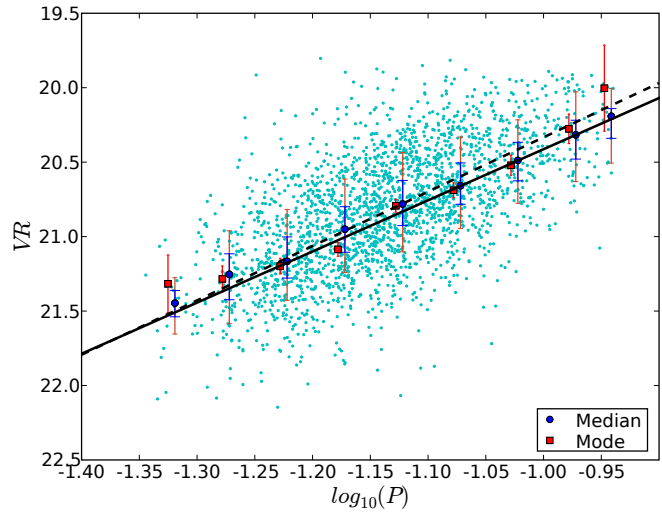}
\caption[]{PL-diagram of all candidates with fitted PL-relation.  The
cyan dots show all \dsctsing~candidates.  The blue circles give the median
magnitudes for each period.  The blue and orange error bars show the
33$^{rd}$ and 66$^{th}$ percentiles of the population respectively,
i.e. 33\% ($\pm$17.5\%) and 66\% ($\pm$33\%) of the candidates within
each bin lie between the ends of the error bars.  We note that the distribution in
each bin is not always symmetric about the median.  The red squares show the
mode for each bin, and their errors show the inverse square root of
the number of sources in the bin.  To improve visibility, we show the
median and mode values slightly offset from the center of the bin
position.  The solid line is the best fit PL relation to the modes
(see Equation~\ref{eqn:PLbest}).  The dashed black line is the best
fit PL relation when fixing the slope to that given by P08 (see
Equation~\ref{eqn:PLp08}).}
\label{fig:dScutiPL_allFit}
\end{figure}

\noindent Using this equation, we fit for the PL relation for the
SuperMACHO \dsct.  Because of the possibility of subgroups, we perform
the fit using the ``ridgeline'' of the PL diagram.  This assumes that
the main population of \dsct~is the most populous.  To find the
"ridgeline" we first bin the data by period (0.05~day binsize) and
then by \VR~(20 bins for each period bin).  We then take the densest
\VR~bin for each period bin as the mode.
Figure~\ref{fig:dScutiPL_allFit} shows the median and mode magnitudes
for each bin.  Using the mode as the dependent variable and the
inverse square root of the number of candidates in each period bin as
the uncertainty, we perform a linear, least-squares fit to the data.
This yields the PL relation:

\begin{equation}
\VR = -3.65 \log_{10}P + 16.68\pm0.11~{\rm mag}
\label{eqn:PLp08}
\end{equation}

We also independently determine the slope of the PL relation using the
SuperMACHO \dsct.  We again perform a fit to the ridgeline, this
time leaving the slope as a variable parameter.  This fit yields the
relation:

\begin{equation}
\VR = -3.43\pm0.26 \log_{10}P + 16.98\pm0.30
\label{eqn:PLbest}
\end{equation}

\noindent Similar to Equation~\ref{eqn:vrint}, we determine the
intercept, $\VR_{c,-3.43}$, for all candidates using the slope from
the best fit.  Figure~\ref{fig:VRintHist} shows the histogram of
$\VR_{c,-3.43}$ (hereafter, $\VR_{c}$ refers to the
PL-corrected-luminosity using this best fit slope).  We examine the
impact of our binsize on the best fit to the ridgeline.  We find that
both doubling and halving the binsize give shallower slopes.  In both
instances, however, this is because the selected binsize gives too
much weight to the sparse region of the PL-diagram at short periods. 

\subsection{Multimode pulsators}\label{sec:SpecAn}

\begin{figure}
\centering
\includegraphics[angle=0,width=0.8\linewidth]{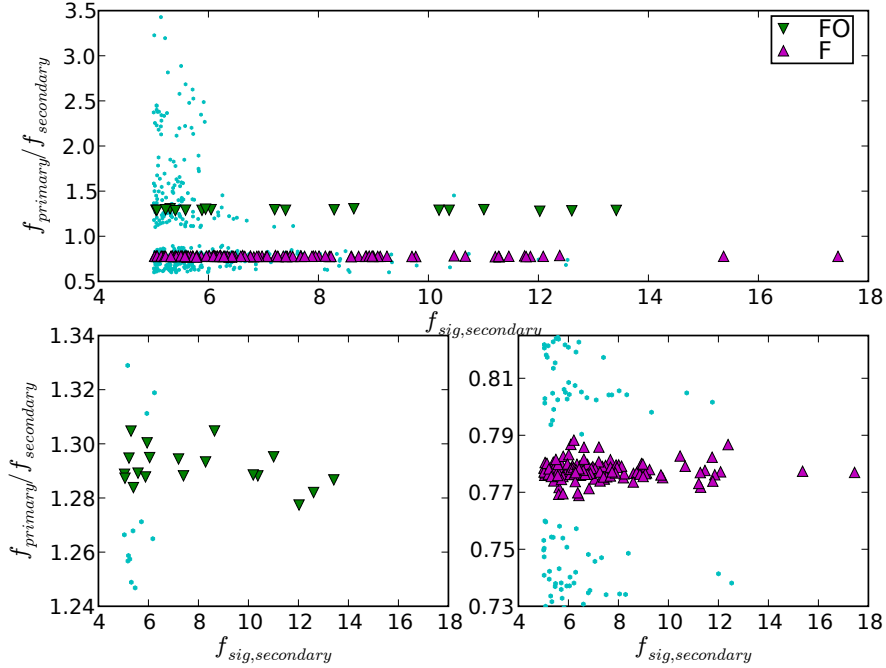}
\caption[]{Ratio between primary and secondary frequencies against
  $f_{\rm sig,secondary}$.  We plot the ratio of the primary to the
  secondary frequency of variability against the significance of the
  secondary frequency for all candidates showing secondary frequencies
  with significance greater than 5 \citep[see][]{SigSpec07}.  We have
  removed any candidates showing only secondary frequencies that have
  whole number ratios to the primary.  We find 119 candidates with
  $0.765 < f_{\rm primary}/f_{\rm secondary} < 0.790$ (green
  pentagons) and 19 candidates with $1.275 < f_{\rm primary}/f_{\rm
    secondary} < 1.305$ (maroon diamonds).  These likely correspond to
  double-mode and overtone pulsators respectively.  We note the
  appearance of less dense additional clusters of candidates with
  secondary modes at frequency ratios of $\sim0.74$ and $\sim0.8$, and
  we identify an outlier at $f_{\rm primary}/f_{\rm secondary} =
  1.452$.  The close proximity to of these clusters to ratios with the
  first overtone frequency suggest that these may also belong to the
  groups described above.  As shown in Figure~\ref{fig:MMnoGroup},
  these light curves show significantly less scatter with respect to
  their characteristic photometric uncertainties than those falling
  into the above classifications, even for similar values of $f_{\rm
    sig,secondary}$.  We conclude that the secondary frequency
  determination for these sources may be unreliable.  }
\label{fig:ratvfsig2nd}
\end{figure}

\begin{figure}
\centering
\includegraphics[angle=0,width=0.6\linewidth]{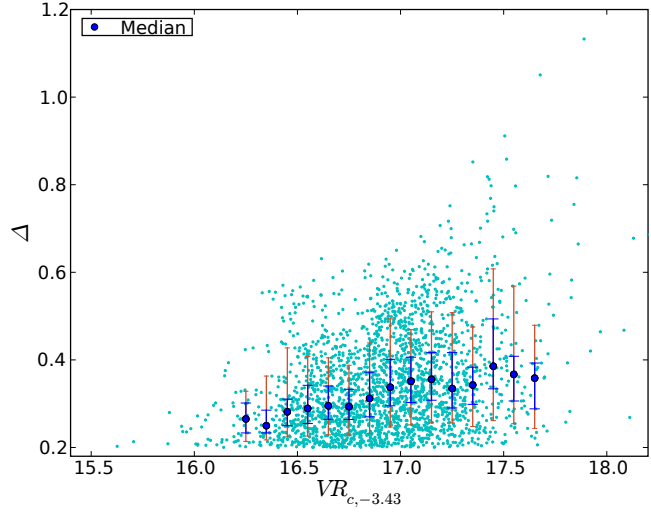}
\caption[]{$\Delta$ against $\VR_{c}$ for all candidates.  $\Delta$
gives the peak-to-trough amplitude of the phased light curve.  We bin
the candidates by $\VR_{c}$.  The blue circles show the median
amplitudes for each bin.  The error bars show the 33$^{rd}$ and
66$^{th}$ percentile as described in the caption of
Figure~\ref{fig:dScutiPL_allFit}.  We do not show bins containing
fewer than 20 candidates.  We find that the brightest candidates have
the lowest amplitudes.  We also find few candidates having $\Delta >
0.4$~mag that are brighter than $\VR_c = 16.8$.}
\label{fig:amplitudevInt}
\end{figure}

In addition to the primary frequency of variation, SigSpec finds lower
amplitude modes of variation.  We use the SigSpec frequency spectrum
to determine whether the candidates show multiple modes of pulsation.
We consider any frequency within $\frac{1}{10}$ of a whole number
ratio to the primary frequency to be an alias.  We only consider
secondary frequencies with a spectral significance greater than 5
\citep[see][]{SigSpec07}.  SigSpec does return additional modes with
significance greater than 5 for a few candidates.  Given the typical
signal-to-noise ratio and sampling of the SuperMACHO data, however, we
do not consider these to be reliable.\footnote{As described in Section~\ref{section:diffphot} \citet{Garg08} finds
that the typical systematic error in SuperMACHO difference flux
measurements is 0.01~mag, and characteristic uncertainties for faint
sources are a few percent.  The median pulsation amplitude of the
\dsct~is approximately 0.3~mag (see Figure~\ref{fig:amplitudevInt}),
suggesting that typical measurement uncertainties for the faintest
sources may be up to 10\% of the pulsation amplitude.  We also lack
sufficient coverage to adequately characterize the properties of many
higher frequencies.  For example, \citet{Poretti09} have 140,016
datapoints, while our typical light curve has $\sim$130.  Because of
the limited coverage and relatively large errors in the light curves,
we opt for a conservative cut-off at secondary modes in our multimode
analysis.}  Figure~\ref{fig:ratvfsig2nd} shows the ratio between the
primary and secondary frequency, $f_{\rm primary}/f_{\rm secondary}$,
as a function of the spectral significance of the secondary.
Table~\ref{tab:dScuti_par} gives $f_{\rm primary}/f_{\rm secondary}$
for candidates showing a secondary frequency.

\begin{figure}
\centering
\plottwo{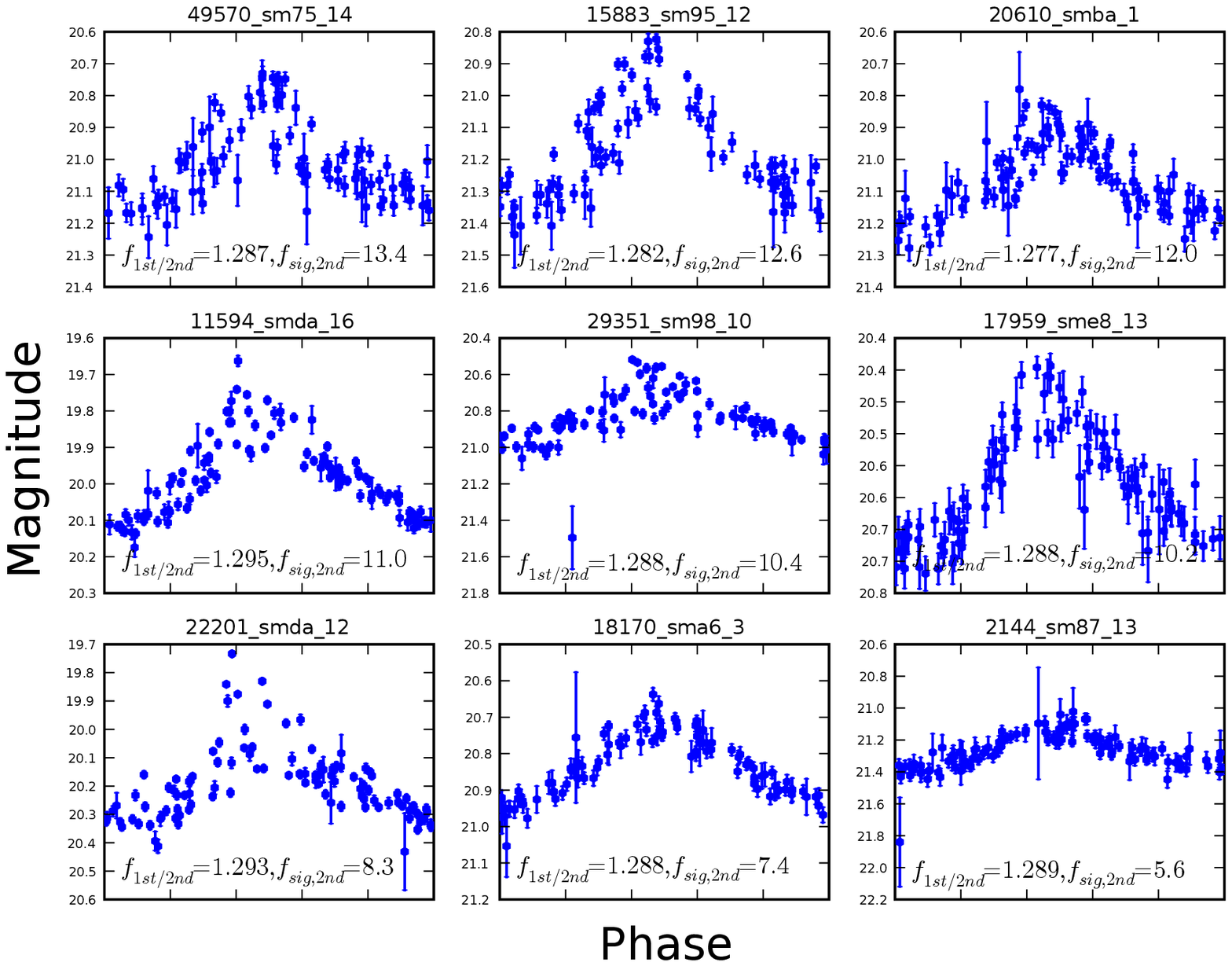}{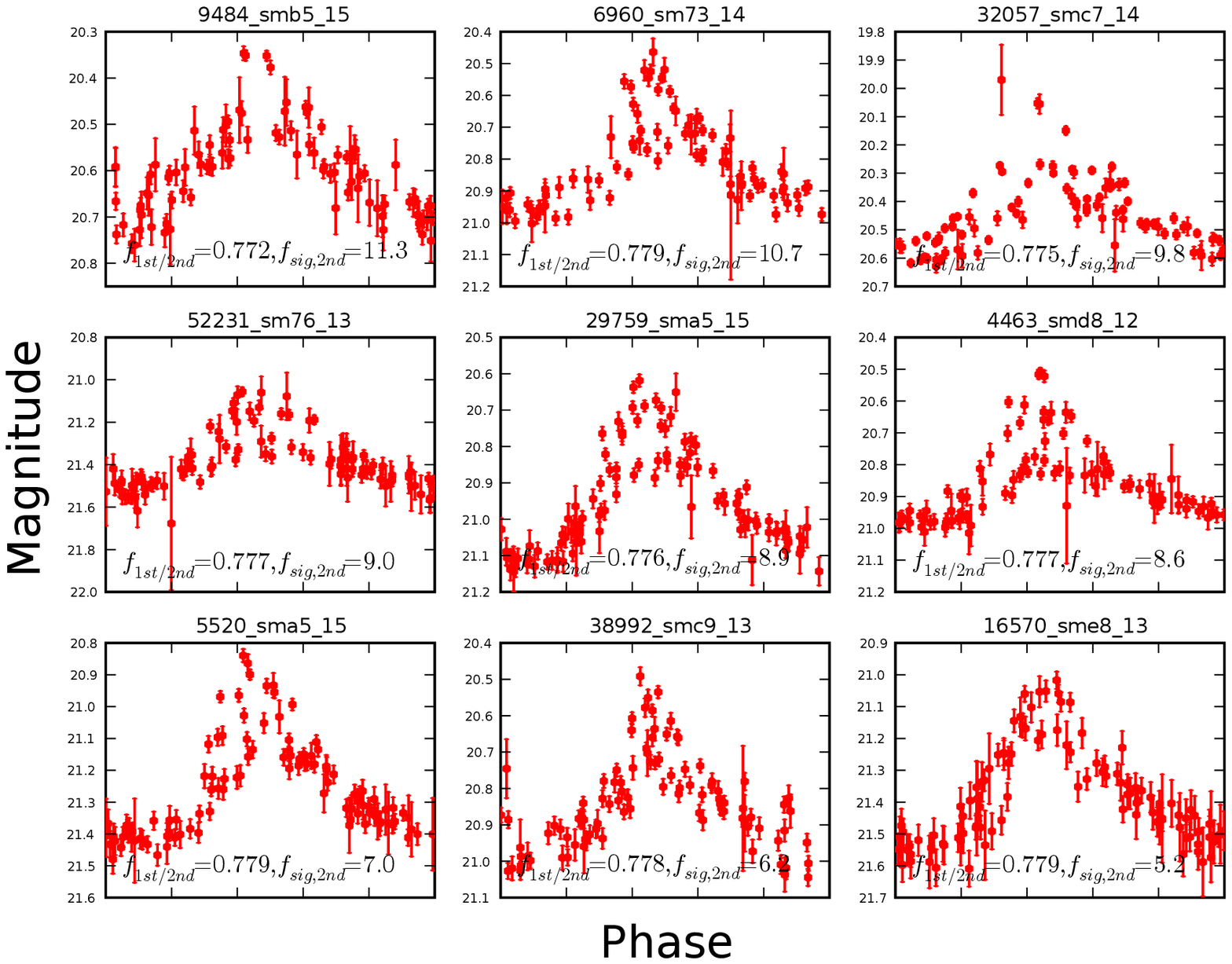}
\caption[]{Light curves for a selection of candidates showing secondary
periods with $f_{\rm sig} > 5.0$.  The plot on the left shows candidates
with $1.275 < f_{\rm primary}/f_{\rm secondary} < 1.305$.  The plot on the
right shows candidates with $0.765 < f_{\rm primary}/f_{\rm secondary} <
0.790$.  These groups correspond to multimode first overtone and fundamental
pulsators respectively.  }
\label{fig:MMLCs}
\end{figure}

\begin{figure}
\centering
\includegraphics[angle=0,width=0.6\linewidth]{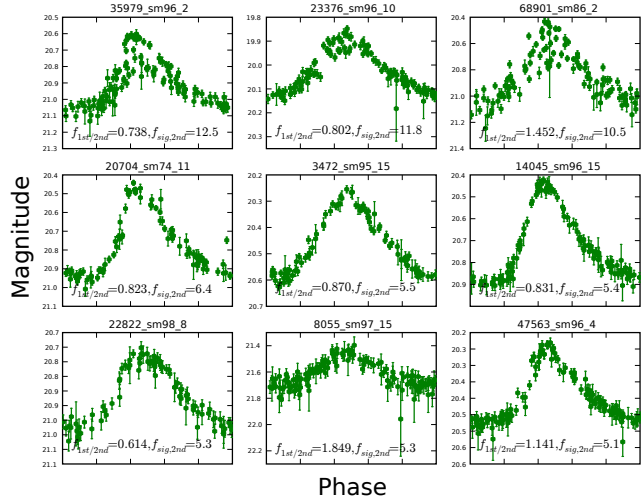}
\caption[]{Light curves for a selection of candidates showing
  secondary periods with $f_{\rm sig} > 5.0$. The ratio $f_{\rm
    primary}/f_{\rm secondary}$ does not fall into the range for
  either F or FO pulsators.  These candidates also show somewhat less
  scatter with respect to their characteristic photometric
  uncertainties.  This suggests that the pulsation amplitudes for both
  modes may be similar which may contribute error to the secondary
  frequency determination.  }
\label{fig:MMnoGroup}
\end{figure}

Examination of Figure~\ref{fig:ratvfsig2nd} reveals two distinct
groups with high values of $f_{\rm sig}$.  We find 119 candidates in the
first group with $0.765 < f_{\rm primary}/f_{\rm secondary} < 0.790$, and we
find 19 candidates in the second group with $1.275 <
f_{\rm primary}/f_{\rm secondary} < 1.305$ (Figure~\ref{fig:MMLCs} shows a
representative selection of light curves for both groups).  The
frequency ratios indicate that these groups are multimode fundamental (F)
and first overtone (FO) pulsators respectively \citep[see][and references
therein]{mcnamaraLMC, Poretti05}.  We note that there are several
other light curves with $f_{\rm sig,secondary}$ greater than 5.0.
Inspection of their light curves reveals that while they do show
additional scatter, it is not as pronounced as that of the candidates
falling into the groups described above (Figure~\ref{fig:MMnoGroup}
shows a representative selection of these light curves).  This may
contribute to greater uncertainty in determining the secondary frequency.

\begin{figure}
\centering
\includegraphics[angle=0,width=0.6\linewidth]{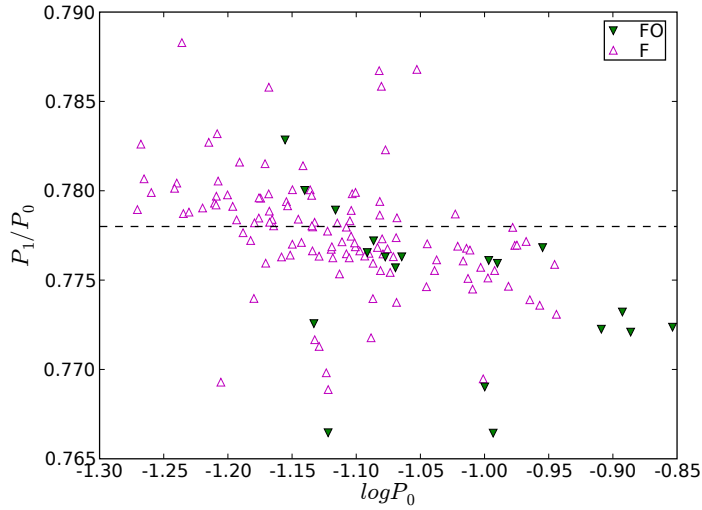}
\caption[]{Petersen diagram.  We show the Petersen diagram for
  multimode F (open maroon diamonds) and FO (green pentagons)
  candidates.  We find several candidates with $P_1/P_0$ greater than
  0.778 (shown as a dashed line).  These values are higher than those
  observed by \citet{Poretti05}.  Higher ratios may be indicative of
  lower metallicity stars.  \citet{Suarez06} also find that higher
  rotational velocities can yield higher ratios; however, their work
  suggests that the impact of rotation is reduced at lower
  metallicities and shorter periods.  }
\label{fig:Petersen}
\end{figure}

Figure~\ref{fig:Petersen} shows the Petersen diagram for the multimode
F and FO candidates.  We have assumed that the
candidates with frequency ratios greater than one are varying in
overtone modes, and set $P_0$ as their secondary period.  We note that
we observe several candidates with ratios above 0.778 whereas
\citet{Poretti05} observe only one.  Based on their findings, this is
likely an indication that these candidates are metal-poor compared to
the Galactic sample.  Such an interpretation is consistent with
observational measurements of LMC metallicity \citep{Cole05}.

We note that the multimode pulsators are concentrated toward fainter
values of $\VR_c$.  We examine whether the lack of multimode pulsators
at the brightest values of $\VR_c$ may be attributable to lower
sensitivity to secondary periods for these candidates.  A plot of
$\Delta$ against $\VR_{c}$ (Figure~\ref{fig:amplitudevInt}) reveals
that the brighter sources generally have smaller amplitudes.  To
determine our ability to detect secondary modes of pulsation, we
simulate several double-mode sinusoidal light curves as they would
appear in our data \citep[see][and Garg et al. 2010, in
prep.]{Garg08,Garg08b}.  We model the light curves using:

\begin{equation}
\VR(t) = A_0 + A_1 \sin(t f_1) + A_1 A_{\rm ratio} \sin(t f_{\rm ratio} f_1)
\end{equation}

\noindent where $\VR(t)$ is the $\VR$ magnitude at time $t$, $A_0$ is
the mean magnitude, $A_1$ is the amplitude of the primary frequency,
$A_{\rm ratio}$ is the ratio between the amplitude for the primary and
secondary frequencies, $f_1$ is the primary frequency, and $f_{\rm
ratio}$ is the ratio between the primary and secondary frequency.  All
parameters except $f_{\rm ratio}$ are randomly selected from a uniform
distribution.  We choose $A_0$ to be between 19 and 22.5.  We choose
$A_1$ between 0.07 and 1.1.  We choose $A_{\rm ratio}$ between 0.1 and
1.0.  We restrict $f_{\rm ratio}$ to be either 0.777 or 1.288,
corresponding to the ratios for multimode F and FO pulsators
respectively.  We note that using our defintion, $\Delta$ is equal to
twice $A_1$ in these simulations.  We find that our ability to detect
secondary modes does not depend strongly on $A_{\rm ratio}$.  For
values of $A_{\rm ratio}$ close to 1.0, we do occasionally misidentify
the secondary period as the primary, but overall our detection
efficiency is roughly 65--80\% for all values of $A_{\rm ratio}$
integrated over $\Delta$.  Using the values of $A_0$ and $f_1$, we
determine $\VR_c$ for each simulated light curve by
Equation~\ref{eqn:PLbest}.  We also find no strong correlation between
$\VR_c$ and our efficiency for detecting secondary modes, though the
fainter light curves tend toward the low end of the efficiency range.
We do find that our efficiency for detecting secondary modes depends
on the amplitude of the primary mode of pulsation, $A_1$.  For $A_1$
close to 0.1, corresponding to $\Delta$ close to 0.2, our detection
efficiency for multimodes in FO pulsators ranges from 15--50\%.  For
multimode F pulsators, it is higher and ranges from 40--60\%.  For no
other parameter do the detection efficiencies differ so greatly
between F and FO frequency ratios.

\subsection{Overtone pulsators}\label{sec:OT}

The histograms shown in Figure~\ref{fig:VRintHist} are heavily skewed
toward brighter magnitudes.  Typically, such a population that lies
above the main PL-relation is thought to be pulsating at the overtone
frequency (see P08 and references therein).  We also find that the
multimode FO pulsators identified in Section~\ref{sec:SpecAn} are
brighter than the multimode F pulsators, though there is no evidence
of secondary pulsation modes in the brightest candidates to help
confirm the FO interpretation.  As discussed
above, we have a low efficiency for detecting multimode overtone
pulsators at small amplitudes.  We note that
Figure~\ref{fig:amplitudevInt} indicates that the brightest values of
$\VR_c$ correspond to the lowest amplitude candidates.  Based on
observations and models of RR~Lyrae, small amplitudes are consistent
with the overtone hypothesis \citep{Bono96}.  As discussed in
Section~\ref{sec:SpecAn}, small amplitudes also reduce our ability to
detect multiple modes particularly in FO pulsators.  While the
presence of multimode FO pulsators amongst the brightest candidates
would lend additional support to the overtone hypothesis, their lack
does not rule out this interpretation.

\begin{figure}
\centering
\plottwo{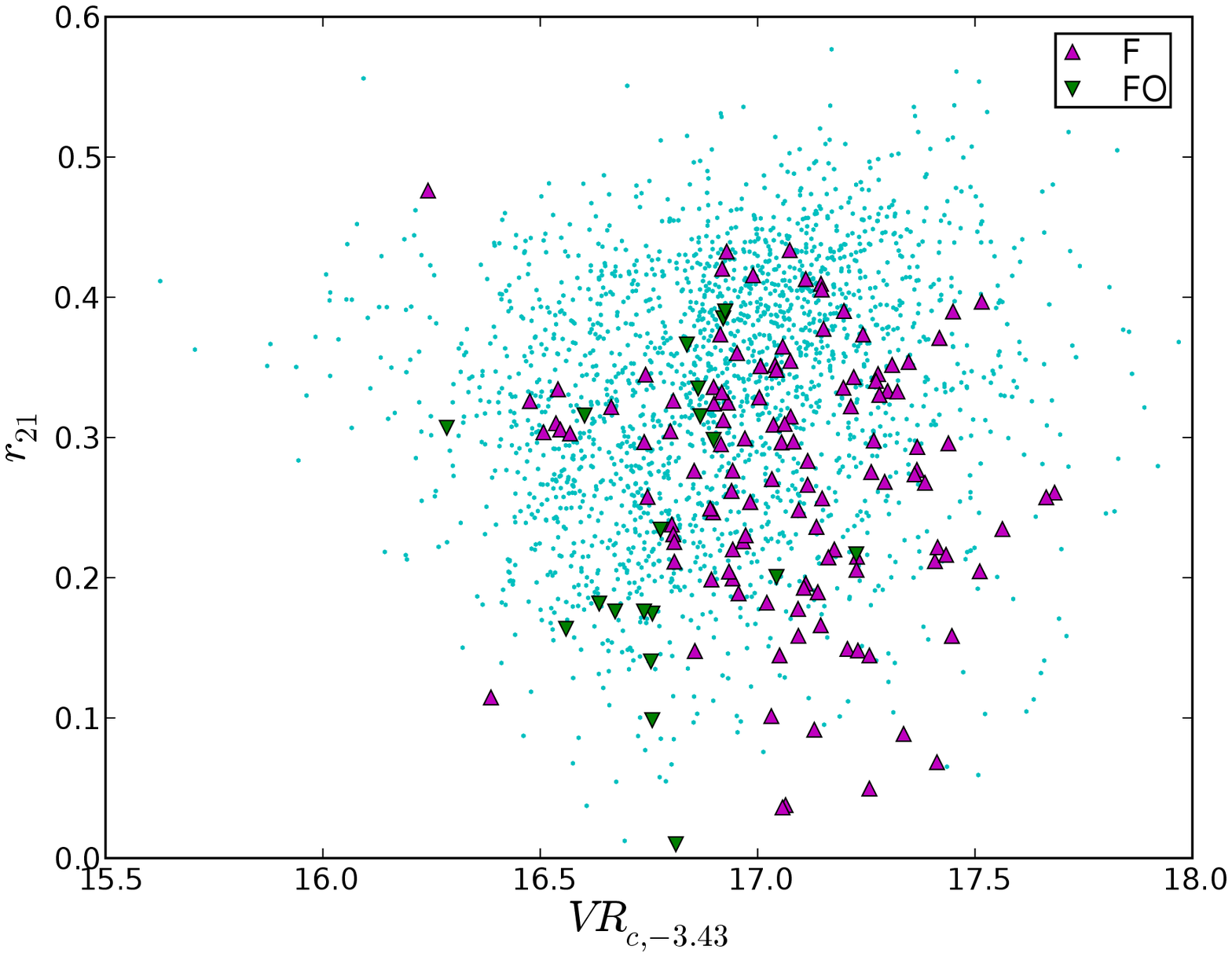}{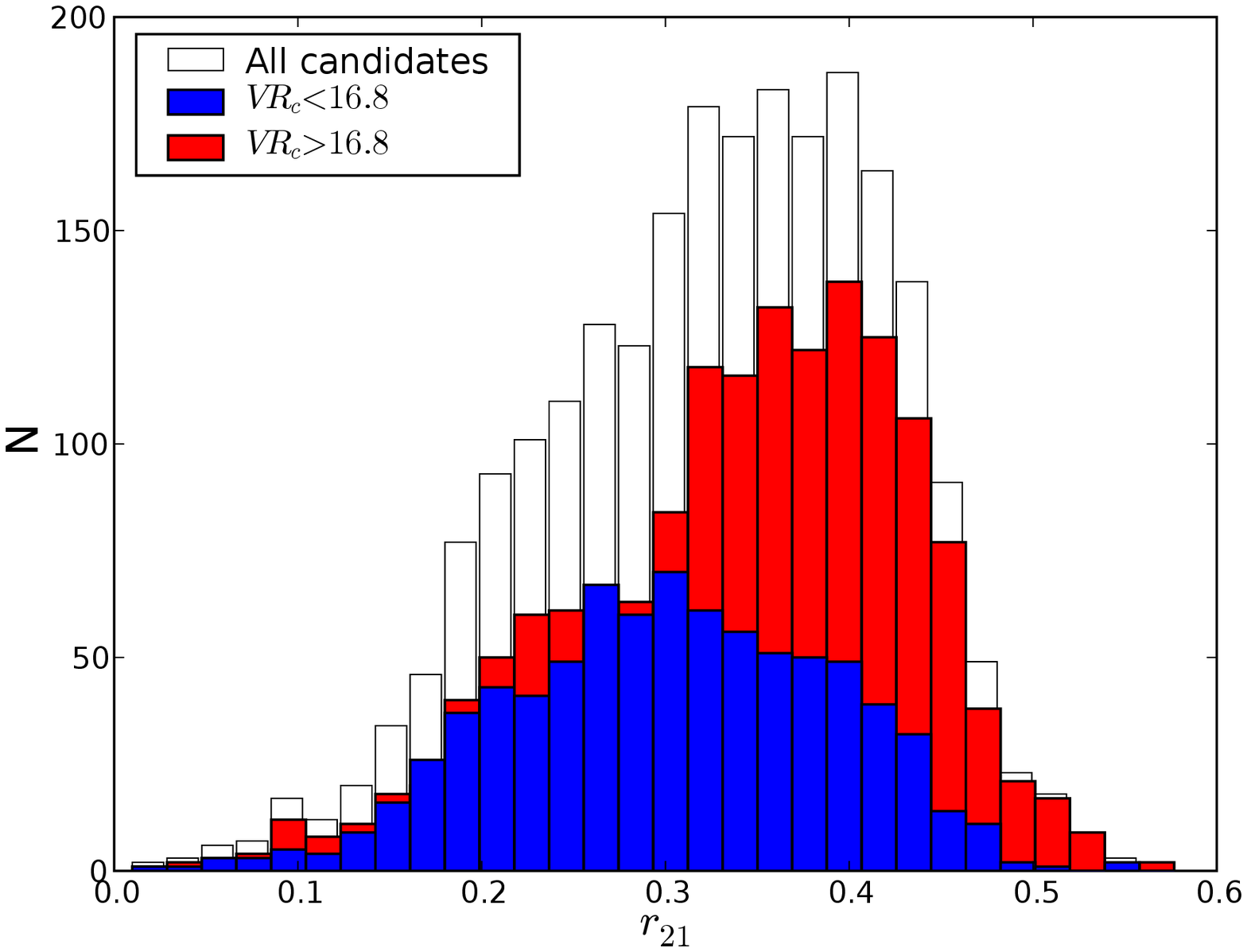}
\caption[]{$r_{21}$ against $\VR_{c}$ and histogram of $r_{21}$ for
all candidates.  Because $\phi_{21}$ is similar for all candidates
(see Figure~\ref{fig:phi21vvrint}), $r_{21}$ measures the relative
contributions of the second and first fourier components to the
overall amplitude.  See caption of Figure~\ref{fig:dScutiPL_all} for
description symbols in the left panel.  In the right panel, white bars
show the histogram of $r_{21}$ for all candidates.  Overploted in red
are fainter candidates with $\VR_c > 16.8$.  Overplotted in blue are
brighter candidates with $\VR_c < 16.8$.  We find that the brighter
values of $\VR_c$ also have smaller values of $r_{21}$ indicating a
more symmetric light curve.  }
\label{fig:r21vvrint}
\end{figure}

\begin{figure}
\centering
\includegraphics[angle=0,width=0.6\linewidth]{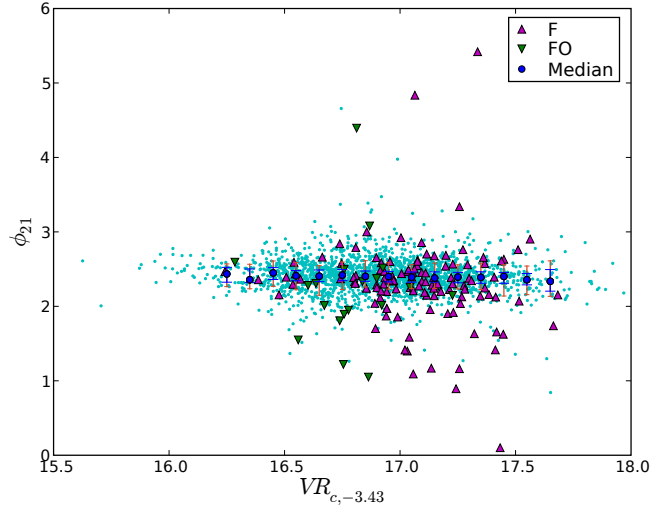}
\caption[]{$\phi_{21}$ against $\VR_{c}$ for all candidates.  We
find that $\phi_{21}$ is similar for all candidates, $r_{21}$ can be
used as a measure of the ratio of the overall amplitud.  Blue circles
and error bars are as described in the caption of
Figure~\ref{fig:dScutiPL_allFit}.  See caption of
Figure~\ref{fig:dScutiPL_all} for description of remaining symbols.  }
\label{fig:phi21vvrint}
\end{figure}

We also examine the shape of the phased light curves using the Fourier
components.  Based on observations of FO RR~Lyrae, we would expect
overtone pulsators to exhibit a more symmetric light curve
\citep{1987ApJ...313L..75S,McNamara00}.  By examining the relation
between $r_{21}$ and $\VR_{c}$ (Figure~\ref{fig:r21vvrint}), we find
this to be the case.  Because $\phi_{21}$ is relatively constant for
all candidates (Figure~\ref{fig:phi21vvrint}), $r_{21}$ measures the
relative contributions of the second and first fourier components to
the overall amplitude.  More asymmetric light curves should have
higher values of $r_{21}$, and we find that lower values of $\VR_c$
correspond to lower values of $r_{21}$.  This is suggestive that the
brighter sources with lower values of $r_{21}$ may be overtone
pulsators having more symmetric light curves.

\begin{figure}
\centering
\includegraphics[angle=0,width=0.6\linewidth]{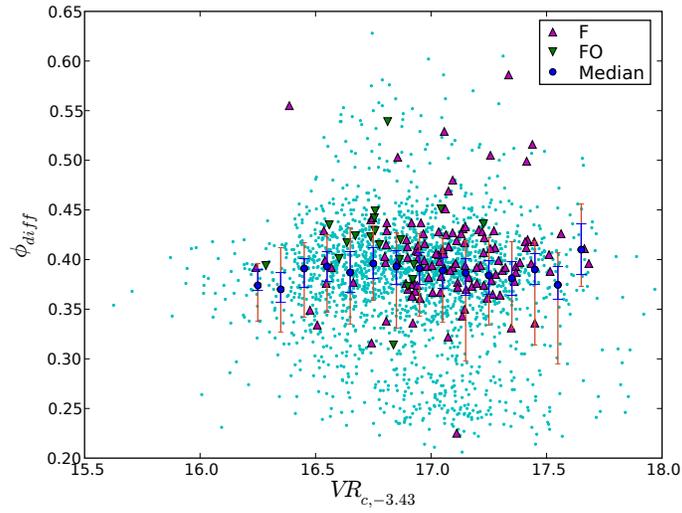}
\caption[]{$\phi_{\rm diff}$ against $\VR_{c}$ for all candidates.
Blue circles and error bars are as described in the caption of
Figure~\ref{fig:dScutiPL_allFit}.  See caption of
Figure~\ref{fig:dScutiPL_all} for description of remaining symbols.  We
find that the lag between $\phi_{\rm max}$ and $\phi_{\rm min}$ is similar for
all candidates, though candidates pulsating in overtone modes tend to
have $\phi_{\rm diff}$ greater than the median.  $\phi_{\rm diff}$ closer to
0.5 indicates a more symmetric light curve.}
\label{fig:phLagvvrint}
\end{figure}

We plot the difference between the phase of minimum and maximum,
$\phi_{\rm diff}$, against $\VR_c$ (Figure~\ref{fig:phLagvvrint}).  For
a purely symmetric, sinusoidal light curve, we would expect
$\phi_{\rm diff}$ to be 0.5.  We find that the majority of our candidates
lie between 0.35 and 0.4.  Notably, however, the FO multimode candidates
have larger values of $\phi_{\rm diff}$
corresponding to a more symmetric light curve as we would
expect.  We find, however, no strong correlation between $\phi_{\rm diff}$ and
$\VR_c$.

By examining the amplitude and shape of the brightest candidates and
comparing to trends observed in RR~Lyrae, these light-curve analyses
indicate that we may have several candidates pulsating in the first
overtone mode.  Notably \cite{Bono97} conclude that the effect may be
reversed for \dsct: overtone \dsct~may exhibit larger amplitudes and
more asymmetric light curves.  Our data disagree with this finding,
however, as the small amplitude candidates examined in this section
lie {\it above} the main PL-relation, indicating that they have shorter
periods than the majority of the candidates with similar luminosities.

We also note that unlike FO RR~Lyrae and FO Cepheids, we see no clear
separation between the F and possible FO \dsct.  Our results are
similar to those of P08 whose PL-diagram also does not show a clear
separation.  We consider whether the lack of separation indicates
significant scatter in the metallicities of the candidates.  Such
scatter would shift the intercept of the PL-relation and potentially
smear the separation between fundamental and overtone pulsators.
\citet{Cole05} do find a significant spread in LMC metallicity based
on spectroscopic observations of the Calcium~II triplet in red giants
toward the bar region.  They find a primary population with [Fe/H]~$=
-0.37$ and $\sigma = 0.15$ and a second, metal-poor population with
[Fe/H]~$= -1.08$ and $\sigma = 0.46$.  Using the metallicity dependence of
0.19~[Fe/H] from \citet{McNamara04}, we might expect an additional
0.03~mag of scatter in the PL-relation based on these results.

Differing extinction may also contribute additional scatter.  If we
assume the redding distribution is $E(B-V) = 0.13$ with $\sigma =
0.045$ \citep{HarrisLMC} and $R_v = 3.1$, we expect an additional
0.14~mag of scatter from variations in extinction.  The tilt of the
LMC will also contribute additional scatter of $\sim$0.01~mag
\citep{2002ApJS..142...71S}.  We examine the uncertainty in our period
determinations as this may also contribute to additional scatter.
Using the technique used for the multimode light-curve simulations
described in Section~\ref{sec:SpecAn}, we simulate \dsctsing~light curves
as they appear in our data.  We use the fourier coefficients from our
candidate \dsct~and Equation~\ref{eqn:expansion} to model the light
curves.  We find that the periods we measure for more than 80\% of the
simulations we recover are within 0.001~days of the input period.  The
additional scatter contributed from this uncertainty is negligible.

We add in quadrature the additional scatter contributed to $\VR_c$ by
the above factors to obtain an expected scatter of 0.14~mag.
Inspection of Figure~\ref{fig:VRintHist}, however, reveals that the
difference between the mode of $\VR_{c,-3.43}$, 17.0~mag, and the
next-densest bright bin, 16.7~mag, is 0.3~mag.  This difference is
more than 2 times the additional scatter, large enough that it is
difficult to explain the lack of a clear separation between F and FO
pulsators based on these factors alone.  These data may indicate that
a clear separation between the PL-relation for F and FO modes does not
exist for \dsct.  This is not necessarily surprising.  \dsct~lie in a
region of the instability strip that spans a wide range of
temperatures and luminosities.  Because of this we would expect the
PL-diagram to have a broad intrinsic width resulting in regions of FO
pulsation that overlap regions of F pulsation.

\subsection{Larger amplitude population}

\begin{figure}
\centering
\includegraphics[angle=0,width=0.6\linewidth]{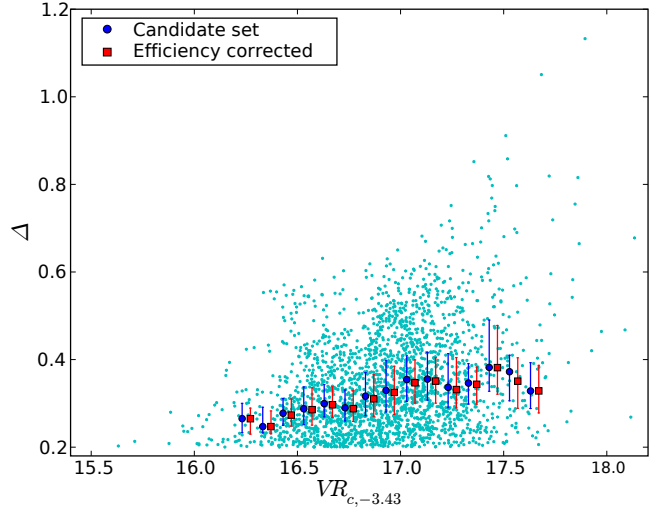}
\caption[]{Efficiency corrected plot of amplitude against $\VR_c$.  We
show a plot similar to Figure~\ref{fig:amplitudevInt}.  The red
squares show the efficiency-corrected median values.  We show the
original (blue circles) and corrected (red squares) 33$^{rd}$
percentile errors (see caption of Figure~\ref{fig:dScutiPL_allFit}).
For visibility, we show these values slightly offset from the center of the bin
positions. }
\label{fig:amplitudevIntcorr}
\end{figure}

The histogram of $\VR_{c}$ shows no strong indication of the
subluminous population described in P08.  We find, however, that the
median $\Delta$ increases for fainter sources
(Figure~\ref{fig:amplitudevInt}).  To test whether this reflects a
selection bias against smaller amplitudes for fainter sources, we use
the simulations of $\delta$-Scuti light curves described in
Section~\ref{sec:OT}.  We find that while we have a relatively high
efficiency for detecting \dsct~at all amplitudes and luminosities
considered ($>$65\%), for the faintest sources we are $\sim13\%$ more
efficient at detecting larger amplitude (0.8~mag $< \Delta <$ 1.0~mag)
variables than smaller amplitude (0.2~mag $< \Delta <$ 0.4~mag)
variables.  The ratio between detection efficiencies for different
amplitudes is similar for brighter sources.
Figure~\ref{fig:amplitudevIntcorr} shows the median and 33$^{rd}$
percentile error bars after correcting for detection efficiency.  We
find that the median amplitude still increases with $\VR_{c}$.

\begin{figure}
\centering
\includegraphics[angle=0,width=0.6\linewidth]{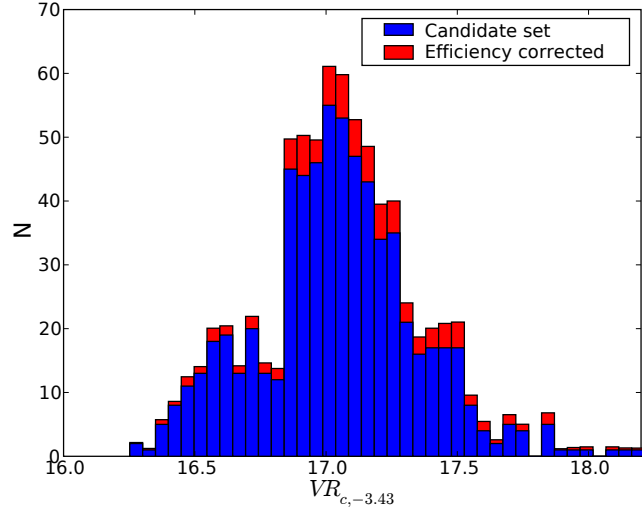}
\caption[]{Efficiency corrected histogram of $\VR_c$ for only sources
with $\Delta > 0.4$.  The blue bars show the observed histogram, and
the red bars give the efficiency-corrected histogram.  We find an
indication of an excess of larger amplitude, subluminous sources
around $\VR_{c} = 17.4$.  This may be the subluminous population
observed in Fornax by \citet{Poretti08}.
}
\label{fig:VRintHistHicorr}
\end{figure}

We also find that the majority of sources with $\Delta >$ 0.4~mag have
$\VR_c$ fainter than 16.8~mag.  We note that the sources in P08 also
have larger amplitudes, generally greater than 0.4~mag.  A histogram
of only the SuperMACHO LMC candidates with $\Delta >$ 0.4~mag also
indicates an excess of subluminous sources similar to that observed by
P08 (Figure~\ref{fig:VRintHistHicorr}).  We suggest
that the subluminous population described in P08 may reflect these
larger amplitude sources.

As discussed in P08, the explanation for these sources remains an open
question.  Observations and models of RR~Lyrae show that for
fundamental pulsators amplitude increases at lower metallicities
\citep{Bono96}.  This may suggest that these candidates represent a
more metal-poor population.  We consider whether they may belong to
the metal-poor LMC population observed by \citet{Cole05} having [Fe/H]
$= -1.08\pm0.46$.  Again using the metallicity dependence from
\citet{McNamara04} of 0.19~[Fe/H], we would expect \dsct~from this
population to lie 0.21~mag below the main PL-relation.  Our data show
a larger separation of almost 0.4~mag.  Notably, the separation
between the main and subluminous populations observed by P08 is
similar.  Given the differences in the compositions and star-formation
histories of Fornax and LMC, this similarity lends support to the
conclusion that these large amplitude, subluminous sources evolved
from the same population as those on the main PL-relation.  It would
otherwise be difficult to explain why both galaxies have such similar
larger amplitude, subluminous populations that evolved independently.

\section{Conclusion}

We have discovered 2323 candidate HADS in the Large Magellanic Cloud
(LMC) using the SuperMACHO set of variables.  Using the SigSpec
software, we performed frequency analyses of these candidates which
reveal several multimode pulsators, including 119 that are pulsating
fundamental modes and 19 in first overtone modes.  We find
evidence for a large set of FO pulsators within this data set.
Notably, the PL-relation defined by the FO pulsators does not show a
clear separation from the PL-relation defined by the F pulsators.
This is not necessarily surprising, as \dsct~occupy a region of the
instability strip that spans a broad region of temperatures, and hence
intrinsic color.  Though we are unable to do so with our single-epoch
multi-band photometry, future surveys that obtain multi-epoch color
information may be able to better define this region of the CMD.  Such
data would also allow for the determination of period-luminosity-color
(PLC) corrected magnitudes that would reveal whether a clear
separation between F and FO pulsators does exist.  

We also find that the majority of our HADS with amplitudes greater
than 0.4~mag lie below the ridgeline of the PL-diagram.  By examining
only these candidates, we find an excess of subluminous sources
similar to that observed in Fornax (P08).  Because the separation
between this population and the main PL-relation is also similar to
that found in Fornax, despite its having different composition and
formation history, we postulate that they form at the same time as
those on the main PL-relation rather than constituting a second, older
and metal-poor population.  Spectroscopic observations to measure the
metallicities of these stars may help to illuminate this discussion.


\section{Acknowledgments}

We would like to thank D.~H.~McNamara for his helpful insights into this
data set.  The SuperMACHO survey was undertaken under the auspices of
the NOAO Survey Program. We are very grateful for the support provided
to the Survey program from the NOAO and the National Science
Foundation. We are particularly indebted to the scientists and staff
at the Cerro Tololo Inter-American Observatory for their assistance in
helping us carry out the survey.  SuperMACHO is supported by the STScI
grant GO-10583.  We are grateful to members of the ESSENCE supernova
survey with whom we work closely.  We would also like to thank the
High Performance Technical Computing staff at Harvard. AG's, KHC's,
MEH's, and SN's work was performed under the auspices of the U.S.
Department of Energy by Lawrence Livermore National Laboratory under
Contract DE-AC52-07NA27344.  C.~Stubbs thanks the the McDonnell
Foundation for its support through a Centennial Fellowship.
C.~Stubbs, AG, and AR are also grateful for support from Harvard
University. AC acknowledges the support of grant P06-045-F
ICM-MIDEPLAN.  DM and AC are supported by grants FONDAP CFA 15010003
and Basal CATA 0609.  LM is supported by grant (CPDR061795/06) from
Padova University.  DLW acknowledges financial support in the form of
a Discovery Grant from the Natural Sciences and Engineering Research
Council of Canada (NSERC).

\clearpage
\begin{deluxetable}{lcccccccccc}
\tabletypesize{\scriptsize}
\tablecaption{$\delta$-Scuti candidates}
\tablewidth{0pt}
\tablehead{
\colhead{ID} & \colhead{RA (J2000)} & \colhead{Dec} & \colhead{{\it Period}} & \colhead{$\Delta$} & \colhead{$V\!R$} & \colhead{$I$} & \colhead{$B-I$} & \colhead{N$_{obs,V\!R}$} & \colhead{$f_{p}/f_s$} & \colhead{Type} \\[-0.30ex]
\colhead{ } & \colhead{ } & \colhead{ } & \colhead{(days)} & \colhead{(mag)} & \colhead{ } & \colhead{ } & \colhead{ } & \colhead{ } & \colhead{ } & \colhead{ } 

}
\startdata
2291\_sm43\_10 & 06:01:06.66 & -72:02:49.8 & 0.10064 & 0.33 & 20.20 & 19.92 & 0.99 & 92 & - & - \\
7345\_sm43\_10 & 05:58:57.54 & -72:02:40.0 & 0.05715 & 0.44 & 21.32 & 21.01 & 0.49 & 92 & - & - \\
2349\_sm44\_10 & 05:59:40.46 & -71:27:14.3 & 0.05988 & 0.27 & 21.12 & 20.74 & 0.65 & 86 & - & - \\
6074\_sm44\_10 & 05:57:58.41 & -71:26:43.9 & 0.05923 & 0.45 & 21.24 & 20.67 & 0.50 & 86 & - & - \\
4223\_sm45\_10 & 05:57:47.82 & -70:52:19.3 & 0.07532 & 0.29 & 21.08 & 20.87 & 1.04 & 82 & - & - \\
4927\_sm45\_10 & 05:57:22.56 & -70:51:14.5 & 0.05847 & 0.24 & 21.32 & 21.09 & 0.97 & 82 & - & - \\
3263\_sm46\_10 & 05:57:46.09 & -70:15:27.2 & 0.06759 & 0.54 & 20.40 & 20.24 & 0.80 & 90 & - & - \\
11884\_sm46\_10 & 05:55:20.03 & -70:14:45.3 & 0.05115 & 0.55 & 20.76 & 20.26 & 0.73 & 90 & - & - \\
7207\_sm53\_10 & 05:51:39.46 & -72:04:59.6 & 0.05647 & 0.38 & 21.38 & 21.00 & 0.42 & 95 & - & - \\
12190\_sm54\_10 & 05:50:15.01 & -71:25:50.2 & 0.08754 & 0.30 & 20.27 & 20.06 & 0.50 & 83 & - & - \\
18598\_sm55\_10 & 05:50:05.53 & -70:52:23.0 & 0.06846 & 0.22 & 20.93 & 20.50 & 0.64 & 80 & 1.400 & double \\
16664\_sm56\_10 & 05:48:30.61 & -70:16:43.6 & 0.08034 & 0.24 & 20.52 & 20.26 & 0.80 & 75 & - & - \\
10335\_sm57\_10 & 05:48:40.49 & -69:40:49.5 & 0.06836 & 0.37 & 20.68 & 20.39 & 0.98 & 92 & - & - \\
12731\_sm58\_10 & 05:47:53.57 & -69:02:16.2 & 0.05311 & 0.43 & 21.47 & 21.14 & 0.87 & 69 & - & - \\
14258\_sm58\_10 & 05:47:40.23 & -69:03:21.2 & 0.07120 & 0.27 & 20.66 & 20.52 & 1.09 & 69 & - & - \\
1859\_sm63\_10 & 05:46:07.28 & -72:02:02.5 & 0.06176 & 0.33 & 21.03 & 20.56 & 0.74 & 90 & - & - \\
2644\_sm64\_10 & 05:45:32.38 & -71:29:03.6 & 0.04764 & 0.36 & 21.34 & 20.98 & 0.82 & 76 & - & - \\
6269\_sm64\_10 & 05:44:44.15 & -71:27:29.0 & 0.08055 & 0.36 & 20.47 & 20.22 & 0.76 & 78 & - & - \\
1783\_sm66\_10 & 05:44:34.91 & -70:15:12.8 & 0.11447 & 0.28 & 20.57 & 20.21 & 1.01 & 88 & - & - \\
14300\_sm66\_10 & 05:43:01.99 & -70:13:31.5 & 0.08786 & 0.22 & 20.19 & 19.83 & 0.90 & 52 & - & - \\
18683\_sm66\_10 & 05:42:33.14 & -70:16:36.4 & 0.07284 & 0.41 & 20.61 & 20.29 & 0.94 & 88 & - & - \\
\enddata
\tablecomments{{\it ID} is the object label for each candidate.  {\it Period} is the primary period of pulsation in days.  $\Delta$ is the overall, peak-to-trough amplitude of $V\!R$-band variation in magnitudes.  N$_{obs,V\!R}$ gives the number of V\!R-band observations.  $f_p/f_s$ is the primary frequency of variation over the secondary frequency of variation.  {\it Type} indicates whether the candidate shows multiple modes of pulsation.  {\it F} and {\it FO} refer to objects identified through their secondary frequencies as double-mode fundamental and first overtone pulsators respectively.}
\label{tab:dScuti}
\end{deluxetable}

\begin{deluxetable}{lcccccccccccccccc}
\tabletypesize{\scriptsize}
\tablecaption{$\delta$-Scuti Fourier Parameters}
\tablewidth{0pt}
\rotate
\tablehead{
\colhead{ID} & \colhead{{\it Period}} & \colhead{$\Delta$} & \colhead{$V\!R$} & \colhead{$I$} & \colhead{$B-I$} & \colhead{$A_1$} & \colhead{$A_2$} & \colhead{$A_3$} & \colhead{$A_4$} & \colhead{$r_{21}$} & \colhead{$r_{31}$} & \colhead{$r_{41}$} & \colhead{$\phi_{21}$} & \colhead{$\phi_{31}$} & \colhead{$\phi_{41}$} & \colhead{$\phi_{\it diff}$} \\[-0.30ex]
\colhead{ } & \colhead{(days)} & \colhead{(mag)} & \colhead{ } & \colhead{ } & \colhead{ } & \colhead{ } & \colhead{ } & \colhead{ } & \colhead{ } & \colhead{ } & \colhead{ } & \colhead{ } & \colhead{ } & \colhead{ } & \colhead{ } & \colhead{ } 

}
\startdata
2291\_sm43\_10 & 0.10064 & 0.33 & 20.20 & 19.92 & 0.99 & 0.14 & 0.04 & 0.01 & 0.00 & 0.29 & 0.10 & 0.03 & 2.53 & 6.22 & 1.93 & 0.42 \\
7345\_sm43\_10 & 0.05715 & 0.44 & 21.32 & 21.01 & 0.49 & 0.20 & 0.05 & 0.02 & 0.01 & 0.26 & 0.10 & 0.03 & 2.53 & 5.45 & 1.16 & 0.43 \\
2349\_sm44\_10 & 0.05988 & 0.27 & 21.12 & 20.74 & 0.65 & 0.12 & 0.04 & 0.02 & 0.00 & 0.36 & 0.13 & 0.01 & 2.53 & 5.55 & 0.99 & 0.43 \\
6074\_sm44\_10 & 0.05923 & 0.45 & 21.24 & 20.67 & 0.50 & 0.18 & 0.06 & 0.02 & 0.02 & 0.32 & 0.09 & 0.10 & 2.33 & 5.72 & 1.95 & 0.39 \\
4223\_sm45\_10 & 0.07532 & 0.29 & 21.08 & 20.87 & 1.04 & 0.12 & 0.05 & 0.02 & 0.01 & 0.43 & 0.17 & 0.11 & 2.45 & 4.93 & 1.39 & 0.37 \\
4927\_sm45\_10 & 0.05847 & 0.24 & 21.32 & 21.09 & 0.97 & 0.11 & 0.03 & 0.01 & 0.00 & 0.23 & 0.10 & 0.02 & 2.77 & 5.60 & 2.55 & 0.47 \\
3263\_sm46\_10 & 0.06759 & 0.54 & 20.40 & 20.24 & 0.80 & 0.23 & 0.07 & 0.01 & 0.02 & 0.33 & 0.06 & 0.07 & 2.40 & 4.83 & 2.20 & 0.36 \\
11884\_sm46\_10 & 0.05115 & 0.55 & 20.76 & 20.26 & 0.73 & 0.25 & 0.09 & 0.03 & 0.01 & 0.36 & 0.13 & 0.04 & 2.51 & 4.90 & 0.85 & 0.36 \\
7207\_sm53\_10 & 0.05647 & 0.38 & 21.38 & 21.00 & 0.42 & 0.17 & 0.05 & 0.03 & 0.01 & 0.28 & 0.16 & 0.04 & 2.69 & 5.14 & 2.68 & 0.52 \\
12190\_sm54\_10 & 0.08754 & 0.30 & 20.27 & 20.06 & 0.50 & 0.13 & 0.05 & 0.01 & 0.01 & 0.39 & 0.06 & 0.05 & 2.37 & 3.99 & 4.06 & 0.33 \\
18598\_sm55\_10 & 0.06846 & 0.22 & 20.93 & 20.50 & 0.64 & 0.09 & 0.02 & 0.02 & 0.01 & 0.20 & 0.19 & 0.15 & 2.90 & 6.06 & 3.00 & 0.43 \\
16664\_sm56\_10 & 0.08034 & 0.24 & 20.52 & 20.26 & 0.80 & 0.11 & 0.03 & 0.00 & 0.00 & 0.28 & 0.03 & 0.03 & 2.37 & 5.06 & 5.20 & 0.38 \\
10335\_sm57\_10 & 0.06836 & 0.37 & 20.68 & 20.39 & 0.98 & 0.15 & 0.04 & 0.02 & 0.02 & 0.27 & 0.15 & 0.14 & 2.53 & 6.08 & 2.31 & 0.41 \\
12731\_sm58\_10 & 0.05311 & 0.43 & 21.47 & 21.14 & 0.87 & 0.18 & 0.05 & 0.02 & 0.02 & 0.28 & 0.11 & 0.10 & 2.53 & 5.30 & 1.76 & 0.40 \\
14258\_sm58\_10 & 0.07120 & 0.27 & 20.66 & 20.52 & 1.09 & 0.12 & 0.03 & 0.01 & 0.01 & 0.23 & 0.07 & 0.10 & 2.47 & 5.63 & 1.74 & 0.40 \\
1859\_sm63\_10 & 0.06176 & 0.33 & 21.03 & 20.56 & 0.74 & 0.15 & 0.04 & 0.01 & 0.00 & 0.30 & 0.04 & 0.03 & 2.56 & 5.98 & 2.15 & 0.39 \\
2644\_sm64\_10 & 0.04764 & 0.36 & 21.34 & 20.98 & 0.82 & 0.15 & 0.06 & 0.01 & 0.01 & 0.43 & 0.04 & 0.08 & 2.49 & 1.06 & 1.84 & 0.37 \\
6269\_sm64\_10 & 0.08055 & 0.36 & 20.47 & 20.22 & 0.76 & 0.14 & 0.06 & 0.01 & 0.02 & 0.41 & 0.10 & 0.11 & 2.21 & 2.97 & 2.82 & 0.34 \\
1783\_sm66\_10 & 0.11447 & 0.28 & 20.57 & 20.21 & 1.01 & 0.12 & 0.04 & 0.01 & 0.01 & 0.29 & 0.09 & 0.07 & 2.07 & 4.56 & 1.48 & 0.37 \\
14300\_sm66\_10 & 0.08786 & 0.22 & 20.19 & 19.83 & 0.90 & 0.10 & 0.03 & 0.01 & 0.01 & 0.29 & 0.06 & 0.06 & 2.17 & 2.90 & 3.85 & 0.36 \\
18683\_sm66\_10 & 0.07284 & 0.41 & 20.61 & 20.29 & 0.94 & 0.18 & 0.04 & 0.02 & 0.01 & 0.22 & 0.13 & 0.06 & 2.67 & 0.09 & 1.01 & 0.45 \\
\enddata
\tablecomments{{\it ID} is the object label for each candidate.  {\it Period} is the primary period of pulsation in days.  $\Delta$ is the overall, peak-to-trough amplitude of $V\!R$-band variation in magnitudes.  $A_n$, $r_{n1}$, $\phi_{n1}$ $\phi_{\it diff}$ are parameters describing the light curves based on a Fourier decomposition (Section~\ref{sec:Fourier}).}
\label{tab:dScuti_par}
\end{deluxetable}

\clearpage

\end{document}